\documentclass[twocolumn]{aastex62}

\submitjournal{ApJL}

\shorttitle{UDG formation mechanism}
\shortauthors{Bennet et al.}

\begin{document}

\title{Evidence for Ultra-Diffuse Galaxy `Formation' Through Galaxy Interactions}

\correspondingauthor{Paul Bennet}
\email{paul.bennet@ttu.edu}

\author{P. Bennet}
\affiliation{Physics \& Astronomy Department, Texas Tech University, Box 41051, Lubbock, TX 79409-1051, USA}
\author{D. J. Sand}
\affiliation{Steward Observatory, University of Arizona, 933 North Cherry Avenue, Rm. N204, Tucson, AZ 85721-0065, USA}
\author{D. Zaritsky}
\affiliation{Steward Observatory, University of Arizona, 933 North Cherry Avenue, Rm. N204, Tucson, AZ 85721-0065, USA}
\author{D. Crnojevi\'c}
\affiliation{Physics \& Astronomy Department, Texas Tech University, Box 41051, Lubbock, TX 79409-1051, USA}
\affiliation{University of Tampa, 401 West Kennedy Boulevard, Tampa, FL 33606, USA}
\author{K. Spekkens}
\affiliation{Department of Physics and Space Science, Royal Military College of Canada P.O. Box 17000, Station Forces Kingston, ON K7K 7B4, Canada}
\affiliation{Department of Physics, Engineering Physics and Astronomy, Queen’s University, Kingston, ON K7L 3N6, Canada}
\author{A. Karunakaran}
\affiliation{Department of Physics, Engineering Physics and Astronomy, Queen’s University, Kingston, ON K7L 3N6, Canada}

\begin{abstract}

We report the discovery of two ultra-diffuse galaxies (UDGs) which show clear evidence for association with tidal material 
and interaction with a larger galaxy halo, found during a search of the Wide portion of the Canada-France-Hawaii Telescope Legacy Survey (CFHTLS).  The two new UDGs, NGC2708-Dw1 and NGC5631-Dw1, are faint ($M_g$=$-$13.7 and $-$11.8 mag), extended ($r_h$=2.60 and 2.15 kpc) and have low central surface brightness ($\mu(g,0)$=24.9 and 27.3 mag arcsec$^{-2}$), while the stellar stream associated with each has a surface brightness $\mu(g)$$\gtrsim$28.2 mag arcsec$^{-2}$.  
These observations provide evidence that the origin of some UDGs may connect to galaxy interactions, either by transforming normal dwarf galaxies by expanding them,  or because UDGs can collapse out of tidal material (i.e. they are tidal dwarf galaxies).
Further work is needed to understand the fraction of the UDG population `formed' through galaxy interactions, and wide field searches for diffuse dwarf galaxies will provide further clues to the origin of these enigmatic stellar systems.

\end{abstract}


\section{Introduction} \label{sec:intro}

The last several years have seen a resurgence of interest in the low surface brightness universe, and in particular the population of so-called ultra-diffuse galaxies \citep[UDGs;][]{vandokkum15}, a term that refers to the largest, lowest surface brightness objects, with half light radii $>$1.5 kpc and central surface brightnesses $>$24 mag arcsec$^{-2}$.  Although UDGs have been discussed in the literature for some time \citep[e.g.][among others]{Sandage84,Caldwell87,Impey88,dalcanton97,Conselice03}, recent work has found hundreds of examples in cluster environments \citep{vandokkum15,koda15,Mihos15,Munoz15,Yagi16,vdBerg16}, along with lower density group \citep{crnojevic16,toloba16a,Merritt16,Roman17,Spekkens18,Cohen18} and field examples \citep{Bellazzini17,Leisman17,Kadowaki17}.

There is considerable debate as to the origin of UDGs, and it is likely that they are a `mixed bag' of populations with multiple origins \citep[e.g.][]{Zaritsky17,Lim18}.  For instance, some UDGs may be `failed galaxies' with Milky Way-like total masses, but with dwarf galaxy stellar masses \citep[e.g.][]{vandokkum15,vandokkum16,trujillo17} while others appear to  simply be the low surface brightness extension of the standard dwarf galaxy population \citep{Beasley16,Sifon18,Amorisco18}.   Most UDGs with metallicity measurements point to a dwarf galaxy origin consistent with their metal poor stellar populations \citep[e.g.][]{Kadowaki17,F18,Pandya18}. Different formation scenarios posit that UDGs have been subject to extreme feedback, which inhibited early star formation \citep[][]{DiCintio17,Chan18}, or that they are the `high-spin' tail of the dwarf galaxy population \citep[][]{Amorisco16}.  A more prosaic explanation would be that UDGs are the product of tidal and/or ram pressure stripping \citep[e.g.][]{Conselice18}, which can remove stars and expand the galaxy's size \citep[e.g.][]{Errani15}; semi-analytic calculations show that this scenario is viable for cluster UDGs \citep{Carleton18}. Similarly, although this has rarely been discussed in the literature \citep[although see][and their discussion of NGC1052-DF2; \citealt{vandokkum18}]{Trujillo18}, some UDGs could plausibly be large, low surface brightness tidal dwarf galaxies (TDGs).  Born during gas-rich galaxy collisions, TDGs should generally be lacking in dark matter and be metal rich in comparison to normal dwarfs of the same luminosity \citep[e.g.][among many others]{Hunsberger96,Duc12}. This could be a way to produce a dark matter free UDG, such as is claimed for NGC1052-DF1 \citep{vandokkum18}, however in that case interpretation is still under extensive discussion and the presence of a GC population \citep{vandokkum18b} is a significant problem for a TDG interpretation.  
Observationally, some TDGs can survive for $\sim$4 Gyr, and have size and surface brightness properties similar to the recently identified UDG class of galaxies \citep{Duc14}.

There is some observational evidence for a UDG `galaxy interaction' formation scenario in the radial alignment of Coma UDGs \citep{Yagi16}, the kinematics of the globular clusters in at least one Virgo UDG \citep{Toloba18}, and in the very elongated UDG associated with NGC~253 \citep[Scl-MM-Dw2;][]{toloba16a}.  Other UDG-like systems also have suggestive features pointing to a recent galaxy interaction \citep[e.g.][]{Rich12,Koch12,Merritt16,Grecopuff}, or even spatial/kinematic substructure that could result from such interactions \citep[e.g. And XIX;][]{Collins13}.  To our knowledge, the only direct observational evidence that UDG-like objects can form from galaxy interactions comes from a) the disrupting dwarf, CenA-MM-Dw3, which has a $r_{\rm half}$=2.5 kpc and $\mu_0$=26.0 mag arcsec$^{-2}$, with clear tidal streams extending over $\sim$60 kpc in the outskirts of the nearby elliptical Centaurus A \citep{crnojevic16} and b) VLSB-A a nucleated Virgo UDG that has clear tidal features, and is possibly associated with M86 \citep{Mihos15}  






Here we present two additional UDGs discovered during a semi-automated, ongoing search for diffuse dwarf galaxies in the Wide portion of the Canada-France-Hawaii-Telescope Legacy Survey (CFHTLS) -- see \citet{bennet17} for initial results around M101, and a description of our algorithm.  Both UDGs show associated stellar streams  connected to a parent galaxy halo, suggesting that they are being shaped by ongoing galaxy interactions.  This further, direct observational evidence that UDGs can be the product of interactions suggests that this is a viable formation channel for this enigmatic galaxy population.  

\section{The Data and UDG Detection} \label{sec:data}

We are searching for diffuse galaxies in the Wide portion of the CFHTLS, concentrating on fields W1, W2 and W3, using an updated version of the semi-automated detection algorithm presented in \citet{bennet17}.  The total area being searched is $\sim$150 deg$^2$.  The CFHTLS data was taken with the $\sim$1$\times$1 deg$^2$ MegaPrime imager \citep{megacam}, with typical exposure times for each field of $\sim$2750 and 2500s in the $g$ and $r$ bands, respectively.
The fields were downloaded directly from the Canadian Astronomy Data Centre, as were the Point Spread Functions (PSFs) for those image stacks, which were used for measuring dwarf structural parameters and simulating injected dwarfs. The construction and calibration of these stacks used the MegaPipe data pipeline \citep{Gwyn08}, and is described in detail by \citet{gwyn12}.

Here we briefly outline our diffuse dwarf detection algorithm, which has been updated slightly from that presented in \citet{bennet17}; the algorithm borrows elements from previous work \citep[e.g.][]{dalcanton97,vdBerg16,davies16}.  All diffuse dwarf detection is done on the $g$-band stacked images from the CFHTLS.  First, bright stars and galaxies are directly masked by matching source positions with the Guide Star Catalog 2.3.2 \citep{lasker08}, and then fainter objects are identified and masked by a call to SExtractor \citep{bertin96}.  This step leaves only very faint objects ($<$3 $\sigma$ above the background), extended galaxy halos and low surface brightness features remaining in the image.  After masking, each image is binned by 150$\times$150 pixels (28$\times$28 arcsec), a spatial scale chosen to maximize the detection of large, diffuse objects while also remaining sensitive to smaller features.  Another round of SExtractor is run to identify objects on the binned images, and all candidates are forwarded for visual inspection via a web interface, where our final diffuse dwarf candidates are selected.

We implant simulated dwarf galaxies directly into our images before performing the search in order to better characterize our detection efficiency.  Simulated dwarfs are injected in batches of ten, randomly placed throughout each image.  Each simulated dwarf has a S\'ersic profile \citep{sersic68} with index $n$=0.5--2.0, and ellipticity 0.0--0.7, randomly chosen, which is representative of past UDG measurements.  Each dwarf is given a $g$-band magnitude  between $g$=16--23 mag, with half light radii in the range $\approx$2--370 arcsec.  This range of parameters spans that of normal galaxies to the ultra-diffuse (although this statement is distance dependent), and allows a true quantification of our detection efficiency.  Roughly speaking, given our current binning scale for the CFHTLS data, we are $\sim$90\% complete down to a central surface brightness of $\approx$28.0 mag arcsec$^{-2}$ for objects that are brighter than $g$$\approx$22 mag -- we will present our complete detection efficiency results in an upcoming work (P. Bennet et al. in preparation).

While still in progress, our search of the CFHTLS Wide fields have uncovered hundreds of diffuse dwarf candidates, dozens of which are likely UDGs.  We will present their demographics in an upcoming work, and compare our results with other wide-field searches \citep[e.g.][]{Greco18}.  Here we present two remarkable UDGs which clearly show signs of interaction and stripping, either of the UDG or the parent halo, likely pointing directly to their formation mechanism.


\section{Results} \label{sec:results}

\begin{figure*}
\begin{center}
 \includegraphics[width=16cm]{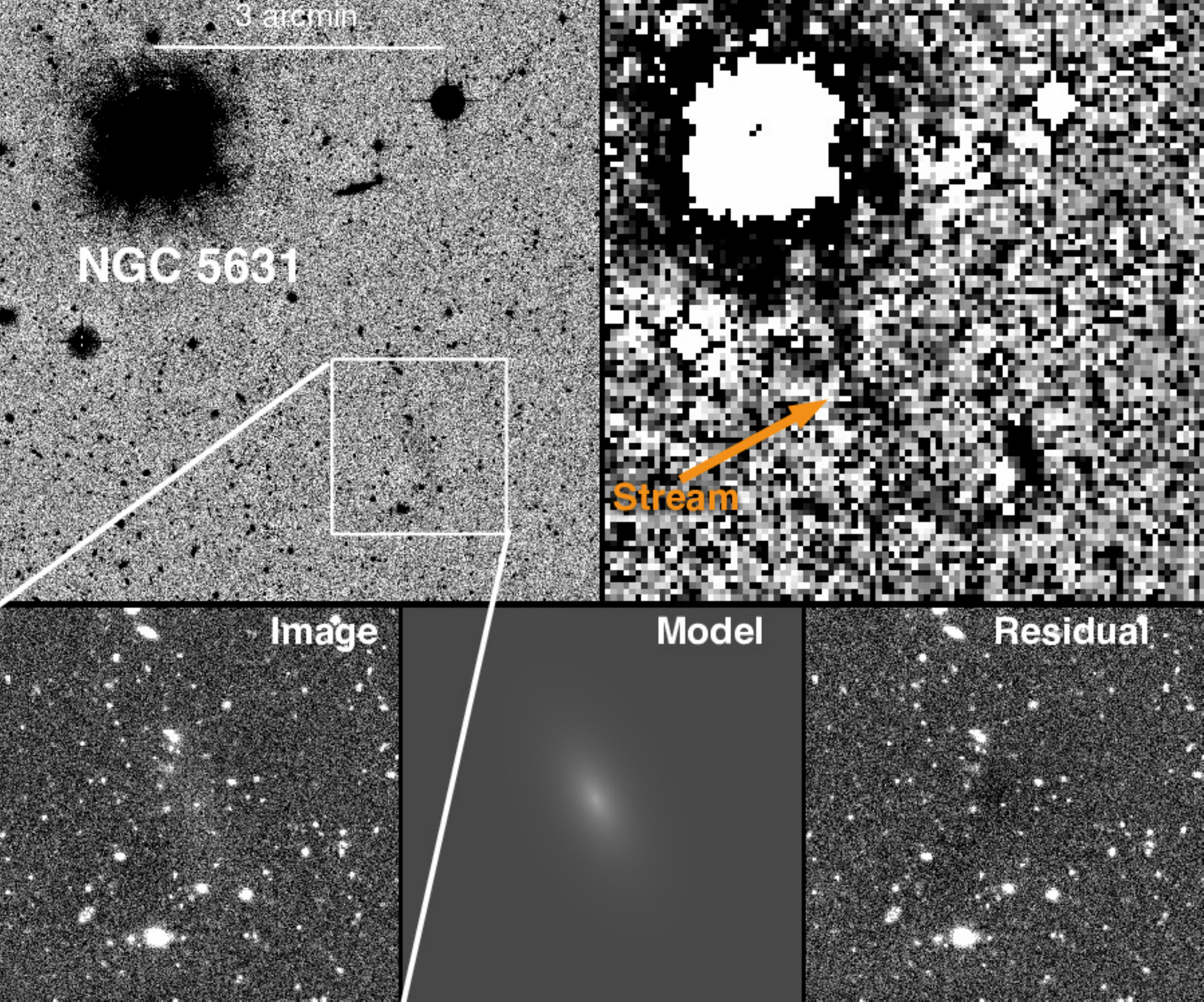}
 \caption{ The $g$-band CFHTLS data of NGC5631-Dw1; North is up and East is to the left in all panels. The upper left image shows the CFHTLS image at full resolution (where the stream is not visible, but NGC5631-Dw1 is just apparent), while the upper right image has been binned and masked to enhance low surface brightness features. The left lower panel shows a zoomed in $g$-band image of NGC5631-Dw1, the center lower panel shows the GALFIT model and the lower right panel shows the residuals.   A clear but very faint stellar stream trails behind the UDG as a dark feature in the binned and masked image.  
  \label{fig:5631}}
 \end{center}
\end{figure*}

During our ongoing search of the CFHTLS, two clear UDG examples exhibited stellar streams connecting them to a parent galaxy halo.  We show NGC5631-Dw1 and NGC2708-Dw1 in Figure~\ref{fig:5631} and \ref{fig:2708}, respectively, including masked and binned versions to highlight the stream associated with each dwarf.  We assume that each object is at the distance of its parent galaxy -- $D$=40.6 and $D$=28.4 Mpc for NGC2708-Dw1 and NGC5631-Dw1, respectively, based on a Tully-Fisher distance for NGC2708 and surface brightness fluctuations for NGC5631 \citep[][respectively]{tully13, Courtois11}.  These distances will have their own associated uncertainty which will effect the inferred physical size and luminosity of each dwarf, although the surface brightness will remain unchanged.

\subsection{Structure \& Luminosity}
The observational parameters for each UDG were derived using GALFIT \citep{peng02}, while the uncertainties were determined by implanting 100 simulated dwarfs with the best-fit properties into our images and re-measuring each with GALFIT; the scatter in these measurements is our quoted uncertainty \citep[see][]{merritt14,bennet17}.  Both objects were fit with a standard S\'ersic profile \citep{sersic68}. We allowed all parameters to vary without restriction for NGC2708-Dw1, but fixed the S\'ersic index to $n$=1 for NGC5631-Dw1 to facilitate the fit, given its extremely low surface brightness.~~
As these objects were very low surface brightness, spatial binning was required. It is also difficult to disentangle the dwarf and its associated stream in the GALFIT process, and there may be an additional systematic uncertainty related to this, although on visual inspection the fits are excellent.  We show our fits and residuals in Figures~\ref{fig:5631} \& \ref{fig:2708}.

\begin{figure*}
\begin{center}
 \includegraphics[width=8.7cm]{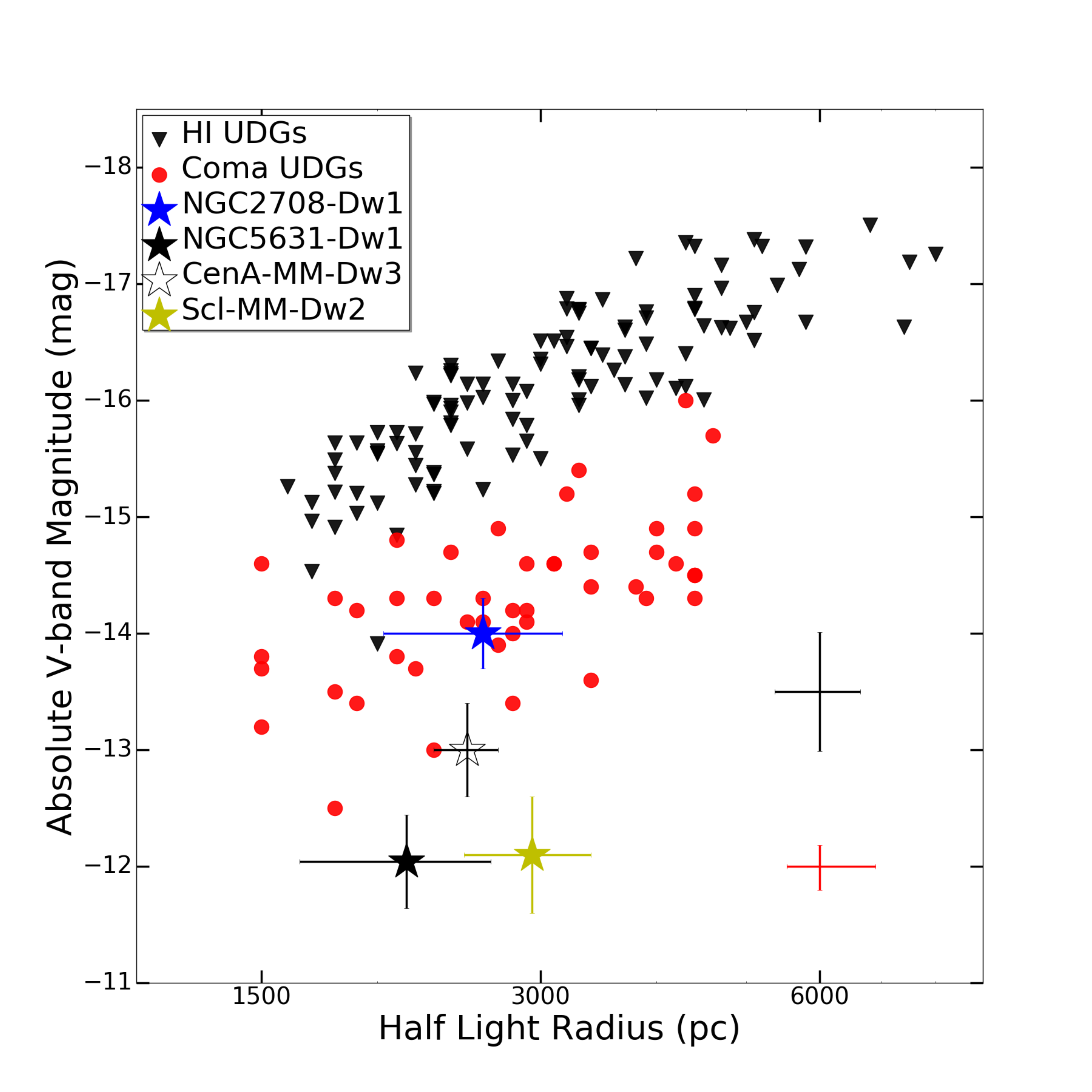}
 \includegraphics[width=8.7cm]{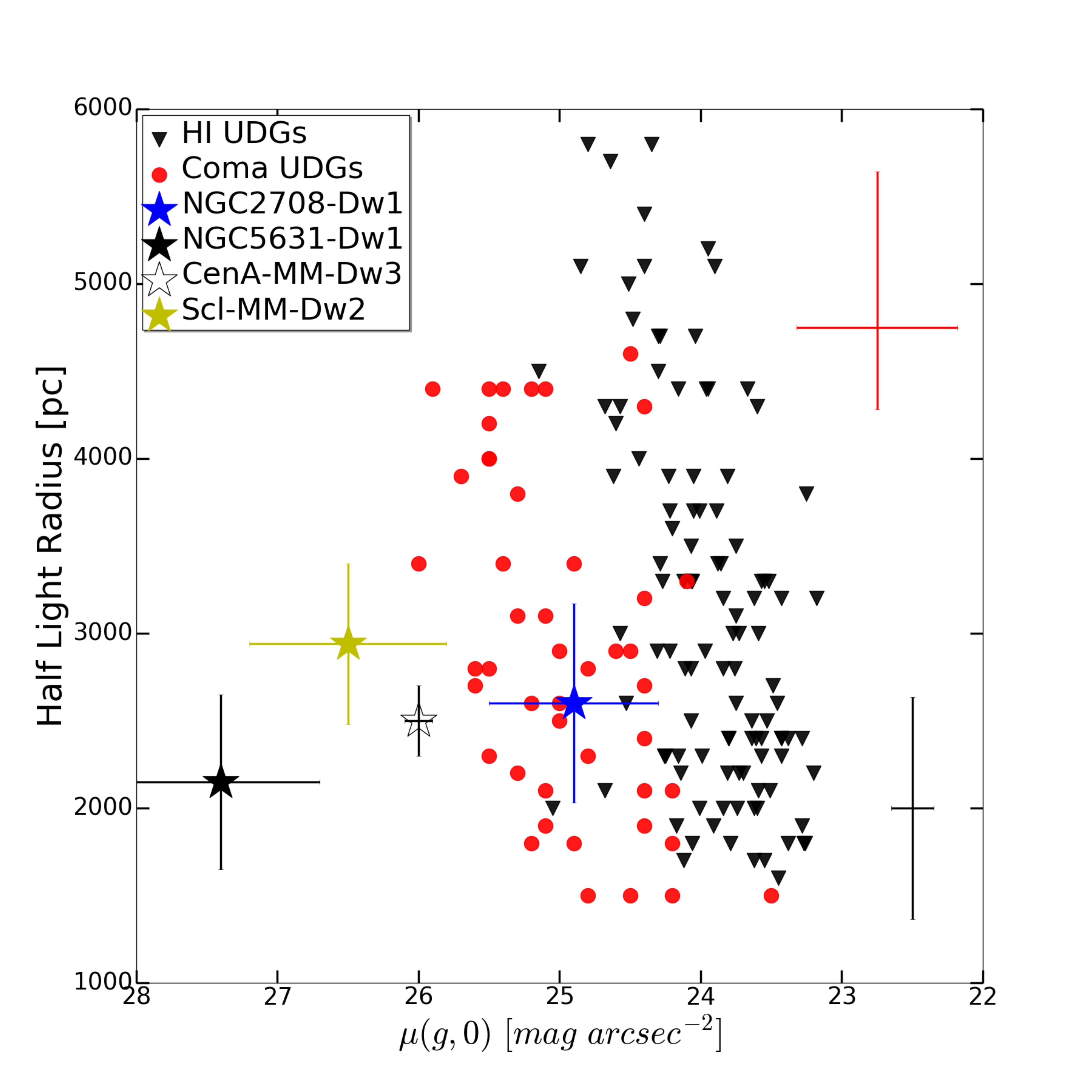}
 \caption{Left -- The size-luminosity relation for NGC2708-Dw1 and NGC5631-Dw1 (stars) compared to other UDG populations --  HI-rich UDGs \citep{Leisman17} are show as inverted black triangles, while the Coma  UDGs (\citealt{vandokkum15}) 
 are shown as red circles. Typical errors for each population are shown on the right of each plot. Also shown are CenA-MM-Dw3 \citep{crnojevic16} and Scl-MM-Dw2 \citep{toloba16a}, the two Local Volume UDGs which show clear signs of disruption. Where direct V-band observations were unavailable they were derived from g and r band data via the procedure in \cite{Jester05}. 
 Right -- The central surface brightness as a function of half light radius for our newly discovered UDGs, plotted with the other UDG populations.     \label{fig:size_lum}}
 \end{center}
\end{figure*}

We put these newly found UDGs in context with those in the literature in Figure~\ref{fig:size_lum}, where we compare them with the UDGs found in Coma \citep{vandokkum15}, and the HI-rich UDG sample of \citet{Leisman17}.  NGC2708-Dw1 has properties which are typical of the general Coma UDG population, with $r_h$=2.60$\pm$0.57 kpc, $M_g$=$-$13.7$\pm$0.3 mag and a central surface brightness of $\mu(g,0)$=24.9$\pm$0.6 mag arcsec$^{-2}$.  NGC5631-Dw1, with a $r_h$=2.15$\pm$0.50 kpc, $M_g$=$-$11.8$\pm$0.4 mag and a central surface brightness of $\mu(g,0)$=27.3$\pm$0.6 mag arcsec$^{-2}$, however, is relatively unique and stands out for its very faint central surface brightness.  Many objects of similarly low surface brightness are found in our general CFHTLS search, and we expect to fill in this surface brightness range in future work.  We also plot the two UDGs in the Local Universe that also show signs of interaction -- Scl-MM-Dw2 \citet{toloba16a} and CenA-MM-Dw3 \citep{crnojevic16}.

We checked {\it Galaxy Evolution Explorer} \citep[GALEX;][]{martin05} imaging at the position of each dwarf, finding no NUV/FUV emission for either object.  From these $\sim$1500 s exposures, we derive NUV $>$20.9 and $>$20.7 mag for NGC2708-Dw1 and NGC5631-Dw1, respectively, and derive a limit on the star formation rate of $\lesssim$3.1$\times$10$^{-3}$ and $\lesssim$3.3$\times$10$^{-3}$ $M_{\odot}$ yr$^{-1}$ \citep{iglesias06} for each object in turn.  The $g-r$ color of the two UDGs are quite uncertain (see Table~\ref{tab:prop}), but given the lack of GALEX detections for each object, they are likely passively evolving at the present epoch (see also the brief HI discussion below).

We estimate the average surface brightness of the streams associated with NGC2708-Dw1 and NGC5631-Dw1 by taking a polygon over the stream area, and aggressively masking intervening, bright sources.  The NGC2708-Dw1 stream is at $\mu$($g$)$\sim$28.2 mag arcsec$^{-2}$, while that of NGC5631-Dw1 is $\mu$($g$)$\sim$28.4 mag arcsec$^{-2}$.  These streams are extremely faint, and may be why similar structures are not more routinely seen around UDGs.

\begin{deluxetable*}{c|cc}
\tablecaption{Stripped UDG Properties \label{tab:prop}}
\tablehead{
\colhead{Name} & \colhead{NGC2708-Dw1} & \colhead{NGC5631-Dw1}
}
\startdata
RA (J2000) & 08:56:12.7 & 14:26:13.6  \\
DEC (J2000) & -03:25:14.8 & +56:31:50.2 \\
$m_{g}$ (mag) & 19.3$\pm$0.3 & 20.5$\pm$0.4 \\
$M_{g}$ (mag) & -13.7$\pm$0.3 & -11.8$\pm$0.4 \\
Color (g-r) & 0.5$\pm$0.4 & 0.4$\pm$0.6 \\
$r_{h}$ (arcsec) & 13.2$\pm$2.9 & 15.6$\pm$3.6 \\
$r_{h}$ (kpc) & 2.60$\pm$0.57 & 2.15$\pm$0.50 \\
S\'ersic index & 1.48$\pm$0.15 & 1.00$^a$ \\
Axis Ratio & 0.83$\pm$0.05 & 0.54$\pm$0.09 \\
$\mu(g,0)$ (mag arcsec$^{-2}$) & 24.9$\pm$0.6 & 27.3$\pm$0.7 \\
$D$ (Mpc) & 40.6 & 28.4 \\
Projected distance (kpc) & 45.2 & 34.1 \\
\enddata
\tablenotetext{a}{The S\'ersic index fro NGC5631-Dw1 is fixed to n=1; see the text for details.}
\end{deluxetable*}

\subsection{Environment}

Both NGC2708-Dw1 and NGC5631-Dw1 are found in a group environment, which is conducive to galaxy encounters \citep[e.g.][]{Barnes85}, and may point to the role that groups play in building up the UDG population across halo masses.

NGC~5631 is an elliptical galaxy, and member of a loose group \citep{Geller83,Pisano04} which is also composed of NGC~5667 and NGC~5678, and possibly several other fainter galaxies. The HI study of \citet{Serra12} shows an HI extension to the SW of NGC5631 in the general direction of NGC5631-Dw1, however this stops short of the position of NGC5631-Dw1 and is not aligned with the stream. This lack of HI (with a limit of $M_{HI}$$\lesssim$5$\times$10$^7$ M$_{\odot}$) within the UDG corroborates the GALEX observations, which indicate it is not actively forming stars.

\begin{figure*}
\begin{center}
 \includegraphics[width=16cm]{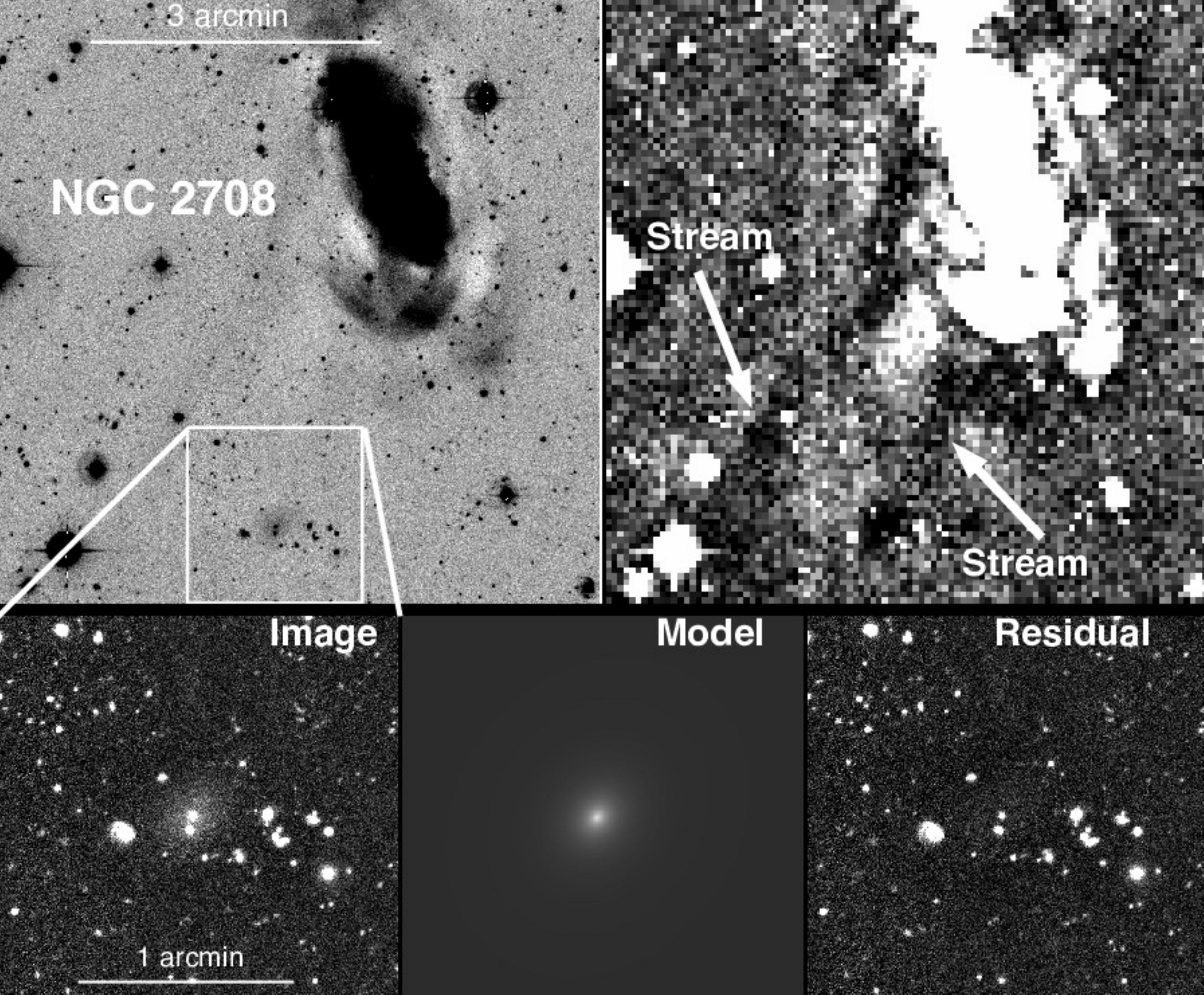}
 \caption{The $g$-band CFHTLS data of NGC2708-Dw1; North is up and East is to the left in all panels. The upper left image shows the CFHTLS image at full resolution, while the upper right image has been binned and masked to enhance low surface brightness features.  Two streams are the dark features apparent in the binned+masked image, one connecting NGC2708-Dw1 to the main body of NGC2708, and another, longer stream just to the north of it. Elements of both streams are also visible in the full resolution image.  The left lower panel shows a zoomed in $g$-band image of NGC2708-Dw1, the center lower panel shows the GALFIT model and the lower right panel shows the residuals.     
 \label{fig:2708}}
 \end{center}
\end{figure*}

The spiral galaxy NGC~2708 is a member of the `NGC~2698 group' as identified by \citet{makarov11}, which has a group velocity dispersion of $\sigma$=94 km s$^{-1}$ and eight identified members.  NGC~2708 itself has undergone several interactions beyond those associated with NGC2708-Dw1.  There is a separate, long tidal stream ($\sim$50 kpc) directly to the north of NGC2708-Dw1 that is visible in Figure~\ref{fig:2708}, which terminates at the same location as a bright foreground star in the southeast portion of the figure.  There is yet another stream which emanates to the north of NGC~2708 (not pictured in Figure~\ref{fig:2708}), approximately 26 kpc long, which also terminates in a fluffy, dwarf-like structure \citep[its morphology is somewhat reminiscent of the `dog leg stream' in NGC~1097;][]{galianni10}.  Portions of this northern stream have been identified previously, and VLA observations reveal it to be HI-rich \citep{pisano02} -- these same HI observations do not show any HI associated with NGC2708-Dw1, with a limit of $M_{HI}$$\lesssim$10$^7$ M$_{\odot}$ \citep{pisano02}, bolstering our argument that this galaxy is no longer forming stars.


\section{Summary \& Conclusions} \label{sec:conc}

We have presented the discovery of two new UDGs with clear evidence for associated stellar streams due to encounters with nearby massive galaxies.  The main body of each dwarf is consistent with the general UDG population (although NGC5631-Dw1 is fainter and lower surface brightness than the bulk of the population), while the stellar streams have estimated surface brightnesses of $\mu(g)$$\gtrsim$28.2 mag arcsec$^{-2}$.  Both UDGs are likely dominated by old, passively evolving stellar populations and reside in a group environment, similar to other (but not all) UDG discoveries.  These stripped objects, along with UDGs discovered in the nearby universe via resolved stellar surveys, point to a possible `formation mechanism' for some fraction of the UDG population.

A scenario where UDGs are produced by galaxy interactions was recently presented by \citet{Carleton18}, and has been suggested elsewhere \citep[e.g.][]{Conselice18}.  In their work, \citet{Carleton18} performed semi-analytic calculations of dwarf galaxies (with both cuspy and cored dark matter halos) in a cluster environment.  Dwarf galaxies with cored dark matter profiles were preferentially shaped by galaxy interactions, causing their stellar mass to decrease and half light radii to increase, and the team was able to reproduce the demographics of the cluster UDG population. It should be noted that \citet{Carleton18} did not recover the observed cluster UDG population with cuspy dwarf galaxy halos, although individual objects did take on UDG-like properties. While these calculations were specifically done for a cluster environment, they should also be applicable to group environments such as that observed in the current work.  

The UDGs in the present work could also be tidal dwarf galaxies (TDGs), the dark matter-free product of gas rich galaxy interactions which continue as cohesive stellar units \citep[for a recent review see][]{Duc12}, and which some observations have shown can be relatively long-lived \citep[$\sim$4 Gyr;][]{Duc14}.  While NGC5631-Dw1 and NGC2708-Dw1 are both associated with the ends of stellar stream material, as might be expected from a TDG scenario, neither has associated HI gas, which seems to be a ubiquitous TDG feature unless the system is very old (although the NGC2708 system appears to have several ongoing encounters, at least one of which is gas rich).  A deep search for neutral gas associated with these UDGs would help clarify their origins. A TDG origin for these objects could also be shown in the mass-metallicity relation; TDGs should be metal-rich compared to equivalent stellar mass dwarf galaxies as they are formed from pre-enriched material from the outskirts of a disk rather than primordial gas \citep{Hunter00}. 

It is also possible that the systems discovered in this work are not long lived structures, and are TDG-like enhancements in tidal streams that match the photometric criteria for UDGs. In this case, it is possible that a portion of the UDG population are chance enhancements of otherwise regular tidal features. 

Additionally, future {\it Hubble Space Telescope} follow-up accounting of the globular cluster (GC) population for these and other UDG systems may also distinguish between formation scenarios -- a TDG origin would have few or no associated star clusters, normal dwarfs would have a few GCs \citep[e.g.][with the caveat that these may be getting stripped in the interactions associated with NGC5631-Dw1 and NGC2708-Dw1]{Zaritsky16}, while more massive UDGs would have commensurately more associated GCs \citep{Beasley16,vandokkum16}.

While local, resolved stellar searches for dwarfs have turned up UDGs that show signs of disturbance \citep{toloba16a,crnojevic16}, direct searches for classical "S"-shaped morphologies among the Coma UDGs have not revealed such tidal features \citep[][although see VLSB-A in the Virgo cluster; \citealt{Mihos15}]{Mowla17}, although the authors admit that they are not sensitive to all  signs of tidal disturbance \citep[see further discussions in][]{Yagi16,Venhola17,Burkert17}.  In any case, it is not clear how long the stellar streams seen in the current work would be visible, as stream lifetimes depend on the dwarf velocity dispersion, stellar radius and orbital eccentricity \citep[see discussion in][]{penarrubia09} -- further modeling of the persistence of tidal features around UDGs in a `galaxy interaction' scenario would help constrain the fraction of the population that forms in this manner.

It is not likely that galaxy interactions can explain the entirety of the UDG population, as an abundant number of field UDGs have been identified \citep[e.g.][]{Leisman17} which have likely never encountered another galaxy.  Note that we also can not rule out a scenario where our UDGs formed by some other mechanism \citep[e.g.][as discussed in Section~\ref{sec:intro}]{vandokkum15,Amorisco16,DiCintio17}, and have subsequently undergone interactions with a larger primary galaxy -- by the same token, one can no longer refute the `galaxy interaction' UDG hypothesis by stating that UDGs show no sign of stripping or interaction.  Future wide-field searches for diffuse dwarf galaxies will reveal their demographics across environments, and hopefully shed light on the origin of the entirety of the UDG population.

\acknowledgments  Research by DJS is supported by NSF grants AST-1821967, 1821987, 1813708 and 1813466.  KS acknowledges support from the Natural Sciences and Engineering Research Council of Canada (NSERC). 
DZ gratefully acknowledges financial support through NSF AST-1713841.
Research by DC is supported by NSF grant AST-1814208.

Based on observations obtained with MegaPrime/MegaCam, a joint project of CFHT and CEA/IRFU, at the Canada-France-Hawaii Telescope (CFHT) which is operated by the National Research Council (NRC) of Canada, the Institut National des Science de l'Univers of the Centre National de la Recherche Scientifique (CNRS) of France, and the University of Hawaii. This work is based in part on data products produced at Terapix available at the Canadian Astronomy Data Centre as part of the Canada-France-Hawaii Telescope Legacy Survey, a collaborative project of NRC and CNRS.


\clearpage

\vspace{5mm}
\facilities{Canada France Hawaii Telescope (Megacam)}

\software{astropy \citep{2013A&A...558A..33A,astropy}, SExtractor \citep{bertin96}, GALFIT \citep{peng02}}
          

\bibliographystyle{aasjournal}
\bibliography{UDG_form_arxiv}

\end{document}